\documentclass{aa}

\begin{document}

\title{Giant Stellar Arcs as the imprint of Precessing
Gamma Jets by   Gamma-Ray Bursts}

\author{Yuri Efremov (1) and Daniele Fargion (2,3)}

\institute{(1) Sternberg Astronomical Institute, MSU, Moscow,
     Russia;\\(2)Physics Dept. Rome, Univ. 1;\\(3) INFN, Rome 1;
      Ple A. Moro 2.Italy}
\date{Received / Accepted}
\authorrunning{Yuri Efremov and Daniele Fargion }
\titlerunning{Giant Stellar Arcs as the Imprint of Precessing
Gamma Jets by  GRBs} \maketitle

\begin{abstract}
Precessing Gamma Jets,  originated by  Neutron Stars or Black
Holes, may blaze to the observer leading to Gamma Bursts (GRBs)
and Soft Gamma repeaters (SGRs). The thin gamma jet is born either
at Supernova (SN) like events mostly at the cosmological
distances, like GRB, or at nearer accreting binary system, like
SGRs. The collimated gamma jet Comptonized by the
ultrarelativistic inner electron pairs jets (hitting thermal
photos) spins (because of the pulsar) and precedes (because of a
companion  or accreating disk) in helical and spiral nebular
shapes. Its cumulative spray may eject and sweep gas creating
nebular rings and plerions and, after long time, partiall HI supershells.
Their consequent gravitational fragmentation may lead to  characteristic
star formation in giant arcs, which  might be just the relic imprints
of the earliest peaked GRBs and the late steady SGRs precesing jet emissions.
Relic stellar rings and arcs may inform, by present morphology on
the jet binary eccentricity and its time evolution. Their lifetime
and occurrence may probe the Supernova-GRB connection.

\end{abstract}

\section{Introduction}

We consider in the present paper  the evidence for strong beaming
of GRB  events. One piece of these data  is the shape of the
possible stellar relics of these events:  the arc-shape stellar
structures were suggested to be the plausible remnants of Gamma
Ray Bursts or related events (Efremov et al. 1998, Efremov
1999a,b). We suggest that  Gamma Ray Burst (GRBs) and Soft Gamma
Repeaters (SGRs) are neither standard candle nor isotropic
(Fireball) explosions. A unified  jet model may explain both of
them as the strong blazing of a light-house, spinning and
precessing gamma jet (Fargion 1998b, 1999). Such jets (born by
black holes (BH) or Neutron Star (NS) in binary or accreting
system) while at their maximal output, as during Supernova like
events ejecting, rarely, in axis  blaze as GRBs. At late, less
powerful but nearer, stages these jets  (as SGRs in our Galaxy and
the LMC) may blaze the observer by similar extreme beaming
($\Omega < 10^{-8}$) and while precessing it may lead to  apparent
variable gamma fluence, respectively comparable, for GRBs, to a
few solar masses annihilation or, for SGRs, to a Supernova
luminosities. Interaction of Gamma-ray bursters (GRB) jets with
the interstellar matter may lead to rapid afterglows tails (in X
and optical bands) as well as to later formation of sweeped gas
supershells, rather similar to those formed by common explosive
SNe, yet generally more energetic and collimated into conical or
hour-glass shapes. (Updated examples of such cumulative spraying
of a variable nebula (Egg Nebula, NGC 2261 ), may be found
respectively easily on net
$(antwrp.gsfc.nasa.gov/apod/image/hst_eggneb_big.gif)$;
$(www.psiaz.com/polakis/n2261/n2261.html)$.The inner Nebula jet
up-down spraying explain the observed twin conical shape and the
evolving nebula luminosity. See also the recent Protostar jet
spraying along HH-34 on net
$(www.eso.org/outreach/press-rel/pr-1999/phot-40b-99-preview.jpg)$.
Such gas ejections at comparable or larger scales are  suggested
here as source for the largest nebular supershells, for which
there is no realistic formation nor after multiple SNe events,
neither by cloud impacts. The samples of such supershells are
mentioned in Efremov et al.(1998), and Loeb and Perna (1998). The
comparison of some parameters of the supershells formed by long
standing supply of energy (like SNe and O-stars) and by single
explosive event were considered by Efremov et al. (1999). However,
some  properties of suggested  stellar remnants of GRBs are
difficult to explain with either mechanism  and  we consider here
the possibility of their formation after action of beamed emission
and/or long-standing multi-precessing jets.

In the present paper we summarize the arguments that the giant
stellar arcs (at least the multiple ones)  did form by the beamed
powerful explosion and/or by the long standing jets from these
objects. The only known such objects are GRBs  and SGRs. If these
giant arcs of stars and clusters are indeed remnants of GRB
explosions (and their long-standing activity), their properties
might give important constrains on the nature of GRB.  The open
angle of arcs being dozens degrees, could point to the beamed at
this angle radiation and the presence of a non-negligible binary
companion. We imagine within the cone jet a more narrow fine
structured jets (arc secs) whose precession angle may be open few
tens of degree and variable because of the disk or companion
interactions or asymmetric accreting disk. We conclude that the
properties of arcs are compatible only with their formation after
the feeding of the narrow multi-precessing jets from GRB. We
believe (Fargion and Efremov;2000) that such multi-precession jet
system is a scale invariant property linking microjets (few solar
masses) to heavier AGN blazars (million to billion solar masses).

\section{The need for beaming}

 The need for GRBs beaming is wide: the GRB luminosities are
 over Eddington, the event peaked structure is chaotic, the spectra
 is non-thermal, the energy budget may exceed two solar masses
 annihilation (Fargion 1994,  Fargion, Salis 1995-98, Fargion 1998-1999).
  The spinning and precessing periodicity is hidden into the
 short GRB observational window;  indeed the periodicity did arise
  in Soft Gamma Repeaters as soon as  more data have been  available.
  As it was demonstrated recently,
 many light curves of the GRB might be explained by the blazing of
multi-precessing   gamma jets (Fargion 1994-1995-1999).
 A wider GRBs data sheet, as  for SGRs data
  would show the spinning periodicity of GRBs and possibly the quasi
  periodic behaviour of the parental  binary system.
 Behind the energy problem stand (to isotropic fireball models)
 the puzzling low probability to observe
any  close GRB as GRB980425 at a negligible cosmic distance (38
Mpc) along with a couple of dozen of far and very far events seen
by  BeppoSax in last two years.

Statistical arguments (Fargion 1998, 1999) favor a unified GRBs
model based on blazing, spinning and precessing thin jet. The far
GRBs are observables at their peak intensities (coincident to SN)
while blazing in axis to us within the thin jet very rarely;
consequently the hit of the target occurs only within a wide
sample of sources found in  a huge cosmic volume. In this frame
work the GRB rate do not differ much from the SN rate. Assuming a
SN-GRB event every 30 years in a galaxy and assuming a thin
 angular cone ($\Omega < 10^{-8}$) the probability to be within
 the cone jet in a ($10^{11}$) galactic sample within
 our present observable Universe volume ($z \leq 2$)
 during a nominal 10 sec GRB
 duration is quite small: ($P < 10^{-3}$). However a precessing
 gamma jet whose decaying scale time (approximated power law $\sim t^{-1}$)
 is nearly  twenty thousand of seconds
 (Fargion 1998-1999) fit naturally the observed  GRB rate.

Also, if these jets have complicated spinning and multi-precession
spirals, they could explain many (or all) features  of the
light-curves of GRB, especially the recent observed periodic tails
in SGR and rarest (20\%) mini-X-GRB precursors (Fargion 2000).
 The possibility that precessing Gamma jets are source by
 their  interactions onto a red giant relic shell
 of the Twin Ring around SN1987A  has been proposed since 1994
 (Fargion \& Salis 1995b, 1995c). It has been also
been suggested (Fargion \& Salis 1995) that the additional
transient presence  of a paraboloid thin arc  along one of the
twin ring of SN1987A,  the mysterious "Napoleon Hat" observed on
1989-1990, was the  evidence for a thin long projected jet
interacting tens parsec away from the SN1987A toward us. The jet
pressure would also accumulate gas and form  dense filamentary
gas.

The possibility of the origin after the GRB explosions was
suggested (Efremov et al. 1998) for the multiple giant arcs of
stars and clusters, described long ago by Hodge (1967) in the LMC
and NGC 6946.  Two of these arcs in the supershell LMC4 region in
the LMC were in a first approximation  explained by Efremov and
Elmegreen (1998) as the results of the multiple SNe near their
centers, yet later on it was demonstrated that their origin after
the GRB explosions is much more probable (Efremov 1998, 1999a,b;
Efremov and Elmegreen, 1998). Here we argue and conclude  that the
most probable cause of the giant arc nature is indebted to the
GRBs and SGRs precessing jets nature.

 A few more stellar arcs are known in other galaxies
 and their open angles, as well as for the LMC arcs,
 are always smaller than $\sim $ 90 degree, what
was suggested to be connected with the beaming of the GRB
explosion (Efremov, 1999), and as we suggest here, it is to be
indebted  to  the driving force of a binary companion.

\section{The giant  stellar arcs}

Giant arcs of the luminous stars and young clusters have been known
for a long time in the region of the supershell LMC4 in the LMC.
The most obvious was noted by Westerlund and Mathewson (1966) and
since then generally called "Constellation III". These authors
ascribed the origin of the arc (which they considered to be the
southern rim of the HI supershell)  to a super-Supernova
explosion. This arc and two others in the same region were
sketched by Hodge (1967), who also found similar arcs in NGC 6946,
recently confirmed by picture given by Larsen and Richtler (1999).
Earlier a number of even larger arcs and rings of clusters were
suspected in different galaxies by Hayward (1964), most of which
are however too large and plausibly just a chance configurations.

The multiple arcs in the LMC contains young clusters of about the
same age (Brown et al., 1998, Efremov and Elmegreen, 1998) and the
same is true for the arcs in the NGC 6946 (Elmegreen et al., 1999).
Along with the strictly circular shape  and giant sizes (150 - 300
pc in radii) these coeval ages  prove beyond any doubt the
coherent origin of objects in arcs. The double small arcs are
suspected in M83 along with two single arcs of clusters and
altogether a dozen of single arcs was noted in galaxies mainly in
Sandage-Bedke atlas  (Efremov, 1999a, 2000b), so the arcs are
rather rare structures. This is possibly related to their short
lifetime as the correlated configurations and also to the
projection effects.

The region of supershell LMC4 was considered the best
manifestation of triggered self-propagated star formation (Dopita
et al. 1985). However, there are serious difficulties  with this
interpretation because there was no evidences for the age gradient
across the supershell (Olsen et al. 1997; Braun et al. 1997).
Recently, Efremov and Elmegreen (1998) suggested that two
well-shaped arcs  in this region formed by triggered star
formation in gas that was swept-up by centralized sources of
pressure. The strictly circular shapes of both arcs are the
strongest evidence for this. Six coeval AI stars near the centre
of the larger arc (called Quadrant, radius  280 pc) were suggested
to be the remnants of an association, including O-stars, which
swepted up the gas in the larger region starting ~30 My ago. A
small cluster near the  center of the smaller arc (Sextant, radius
170 pc) was proposed to be responsible for that one.  The
positions of various features in this region are shown in Figure 1
(where the Quadrant arc in near the center and the Sextant in lower right
corner) and also in Efremov and Elmegreen (1998) and in Figure in
Efremov (1999a).

   These centralized stellar sources of pressure could produce both young
stellar arcs at the right time and position, as Efremov and
Elmegreen (1998) demonstrated, yet the general picture is still
not satisfactory. The main questions remaining are (Efremov,
1999a,b): (1) why are there no giant stellar arcs or rings around
other,  even much more rich, clusters in the LMC,
(2) why are all of the stellar arcs in the LMC  close to each
other and why  are they in only this area,
(3) why are there just arcs and not full stellar rings, and (4)
why are these arcs in the region of the largest and deepest HI
hole  in the LMC?

We believe that points 1 and 2 above are already uncompatible with
assumption that a dozen or a few dozen of SNe in clusters in
centers of the arcs could produce these.  What was so special with
these small clusters? If there was a peculiar initial mass
function, why just for clusters only in this region - more so
because just there exist also the Third arc and probably even one
more, the Fourth arc. We imagine that smaller scale Nebulae and
SNRs whose rings resemble giant stellar arcs are the young
miniature example of large sized GRB and SGRs relics.

There is increasing evidence for general absence inside the HI
supershells of the clusters rich enough to contain the O-stars and
SNe to trigger the formation of supershells. Recently the special
photometric search for clusters, which could produce the
supershells in the irregular galaxy Ho II was carried out by Rhode
et al. (1999). They found  only 6 of total 44 supershells cases
of the presence of clusters (assuming the normal IMF and age)
which could contain SN/O stars numerous enough to form the
observed supershells, These authors stressed that among the most
sure cases of supershells without putative clusters there are just
supershells the most energetic in Ho II; moreover, these
supershells are within the low density regions of the galaxy where
the presence of massive clusters is improbable.

 The recent identification of GRB afterglows (understood as beamed jet tails)
 in distant galaxies
has led to suggestion that they can produce very large shells and
trigger star formation (Efremov, Elmegreen and Hodge, 1998; Perna
and Loeb, 1998, Efremov, Ehlerova and Palous, 1999).  These
suggestion explained the enigmatic supergiant HI shells  without a
central cluster or evidence of an extragalactic cloud impact in
the triggered region. The samples of large energetic supershells
in galaxies with no clusters or evidences for the cloud impacts
are mentioned in papers, referred to above. Also, Rhode et al.
(1999) noted that practically no high velocity clouds exists
around Ho II galaxy which might have been able to form
supershells.

We suggest here that
 supershell may exist also from accumulated jet activity, which probably
 started  with eventual SN birth explosion.

\section{The origin of the GRB progenitors.}

The occurrence of all stellar arcs in the LMC near each other may
be explained  by the common origin of the progenitors of their paternal
GRB  which formed in and then escaped from a massive
near-by cluster. This is compatible with the common assumption
that explosions of some GRBs are the result of mergering of
components of close binaries that include a neutron star or black
hole.  These close binaries might be formed in stellar encounters
inside a dense massive cluster and then escape from it.  There is
indeed such a cluster in the region under consideration (Efremov,
1998, 1999; Efremov and Elmegreen, 1998b). NGC 1978 is within 0.5 - 1.5 kpc
from the arcs and the
center of the LMC4 supershell.  The  age of this cluster  is about
2 Hyr (Bomans  et al. 1995), and it is the richest  cluster of
such an age in the LMC. Its mass is 0.4 - 1.4 millions of suns
(Meylan et al. 1991) and it has a few hundred red giants with
masses of around 1.5 suns.

   The high rate of occurrence of  X-ray binaries (with one component a
neutron star) inside dense globular clusters is well known (e.g.
Bailyn 1996). It was explained long ago as consequence of the high
probability of formation of close binaries after tidal captures in
the dense cluster (e.g. Fabian et al. 1975,  Shklovsky 1982,
Davies 1995, Phinney 1996). It was also shown (McMillan 1986) that
a large number of tidally captured binaries  may escape a dense
old cluster as the result of three-body encounters. The
comprehensive review of the data on binaries and pulsars in
globular clusters (Phinney 1996) lefts no doubts that there is a
lot of possibilities to form close binaries with compact
components in a cluster dense and old enough, and also that many
of these binaries are able  to escape from such a cluster.

There is also a special way for binaries with a neutron star
component to escape from a cluster.  The formation of a neutron
star after a SN explosion in a binary system leads to a high kick
velocity, the most likely value of which is 150 - 200 km/s
(Lipunov et al. 1997). Such high velocities would spread out any
future GRB  over very  large distances around the paternal
cluster. Even much smaller velocities would disperse the GRB
progenitors significantly, because the binaries may take 100 My or
so before they merge to give a GRB (Lipunov et al. 1997).   It is
quite possible that the relics of GRB might be observed from few
to hundreds parsecs from the paternal cluster.

NGC 1978 is also unusual in its extremely flattened shape (Geisler
and Hodge, 1980). This may indicate a formation process involving
the merger of two clusters, especially because no rotation has
been detected (Fisher et al. 1992), or owing to the disk-shocking
and might point to the dynamical state of the cluster which permit
escaping of stars disregarding their masses - and therefore,
binary stars as well.

Many binary system could escaped from this massive cluster and
among them there could be the progenitors of the GRB. Hanson and
Murali (1998) suggested that stellar encounters in globular
clusters were able to produce not only millisecond pulsars but
also binaries that evolve into GRBs.

   The unique stellar arcs, the largest HI hole and the unusual cluster
are not the only peculiar objects in the LMC 4 region. Near NGC
1978  there is excess number of X-ray binary stars, which include
neutron star component and therefore related to GRB progenitors.
Three  X-ray binaries are within 20' of  NGC 1978 and more are in
a wider surrounding, as is evident from Haberl and Pietsch (1999).
The suggested large masses of
these X-ray binaries seems to be uncompatible with them being
escaped from the rather old cluster, yet they might be products
of the complicated evolution inside the dense cluster, including merging
and/or mass exchange.

Anyway, in the same area and close to NGC 1978 is also the object
which is more certain relative to GRB. It is the Soft gamma
repeater (SGR), SGR0526-66, which produced the famous gamma-burst
of March 5, 1979  and is within the most bright SNR in the LMC,
N49. According  Fargion (1998, 1999), Nakamura (1998), Dar (1999),
Spruit (1999) and  other workers,   classical GRB left behind a
soft gamma repeater (SGR) - and the only SGR known in the LMC is
just here!  Also in the same area, at the East end of the Qudrant arc is
the millisecond X-ray pulsar A0538-66, the object of the class which
is considered by Spruit (1999) as  remnants of X-ray binaries that
managed to escape becoming  GRB.
At any rate, whatever could be a reason for this, the
concentration of the giant stellar arcs and GRB progenitors, relatives  and
relics to the same and the only region of the LMC strongly suggest
that GRB and arcs are connected phenomena, and we suggested that
NGC 1978 was the common source of the progenitors of GRB which
produced the stellar arcs and the LMC4 supershell (Efremov
1999a,b).

The escape of the double black holes after close encounters from a dense
cluster need a few Billions years  (Portegies Zwart and McMillan, 1999) and
this is compatible with the NGC 1978 age.  Only after the ejection of
the  binaries with compact components from a cluster they are are close enough
as to go to merging by the gravitational wave emission. This could
explain why there is no X-ray binaries inside the cluster and the stellar
arcs around it and these facts migt be considered the evidences that
the progenitors of GRB are close wbinaries both components of which are
either black hole or neutron stars (Efremov, 2000a,b).

\section{Puzzling arcs properties}

 The properties of stellar arcs, suggested GRB relics, may
say something on the nature of the GRB event. First of all this is
surely the opening angle of the arcs, which is always smaller than
90 degree. The preliminary data on the arcs in a number of
galaxies, including a dozen single arcs point to the preferred
angle of ~60 degree (Efremov, 2000a) yet the real
angle may be much smaller, just because too short arcs are not
recognized as arcs.

Another important property of the arcs is their perfect circular
shape, independent on the galaxy plane inclination (Fig. 1).
This means that some arcs are partial stellar shells, seen in
projection, and not the rings in the plane of galaxy. The outburst
far from the plane of the gas disk of the galaxy should result
in the partial gas shell and owing to the vertical density gradient,
the most dense part of the shell must ve turned to the galaxy plane.
This partial shell  looks like an arc due to the inclination
of the plane  of the galaxy to sky plane (Efremov et al., 1999).

The apex of such a partial shell seen in projection should
always be turned either TO or OFF the line  of nodes of the galaxy plane
(its intersection  with the sky plane).
The most important for us is that this is not the case for the
arcs in the LMC. The apexes of two arcs are about parallel to the
line of nodes, the position angle of which is  162 grades (see Fig. 1).
Therefore, the arcs were NOT formed from the
isotropic outbursts outside the middle of the gas disk of the galaxy.
Therefore, the open angle of the arcs reflects the beaming angle of the
parent explosion, or is the largest possible angle of precession of
the multi-precession narrow jets, which are able to fulfill the
partial shell, forming the swepted-up gas shell and then stellar
spherical partial shell, seen as an arc in projection.

In what follows we give the arguments for the latter possibility:
the precessing jets with the variable angle of the precession. The
idea on spinning-multi-precessing gamma-jets  was advanced already
from the completely other considerations (Fargion 1994, 1995,
1998, 1999).

\section{Gamma Burst and Soft Gamma Repeaters as Multi-precessing Gamma Jets}

It is somehow surprising that after a decade of fireball inflation
papers, at present (GRB990123, GRB990510 and GRB991226 over
energetic event) there is no father or mother spending a word of
regret on the decline and death of their popular isotropic model.
On the contrary there is wide spread resistance to give up this
misleading fireball model.

Gamma Ray Bursts as recent GRB990123 and  GRB990510 emit, for
isotropic explosions, energies as large as  two solar masses
annihilation. These energies are underestimated because of the
neglected role of comparable ejected MeV (Comptel signal)
neutrinos bursts. These extreme power cannot be explained with any
standard spherically symmetric Fireball model. A too heavy black
hole or Star would be unable to coexist with the shortest
millisecond time structure of Gamma ray Burst. Beaming of the
gamma radiation may overcome the energy puzzle. However any mild
''explosive beam'' as some models (Wang \& Wheeler 1998) $(\Omega
> 10^{-2} )$ would not solve the jet containment at the
corresponding disruptive energies. Moreover such a small beaming
would not solve the huge GRBs flux energy windows ($10^{47} \div
10^{54}$ erg/sec), keeping GRB980425 and GRB990123 within the same
GRB framework.

 Only extreme beaming $(\Omega < 10^{-8} )$, by a
slow decaying, but long-lived precessing jet, may coexist with
characteristic Supernova energies, apparent GRBs output and the
puzzling GRB980425 statistics as well as the GRB connection with
older,nearer and weaker SGRs relics. GRBs were understood up to
1998 as isotropic Fireball while SGRs are still commonly described
by isotropic galactic explosions (the Magnetar model). However
early and late Jet models (Fargion 1994-1998,Blackmann et all
1996) for GRBs are getting finally credit. Will be possible to
accept a jet model for GRBs while keeping alive a mini fireball
(based on huge magnetic field energetic budget) for SGRs? Indeed
the strong SGR events (SGR1900+14, SGR1642-21) shared the same
hard spectra of classical GRBs. In particular one should notice
(Fargion 1999a,  1999b,  1999c), the GRB-SGR similar hard spectra,
morphology and temporal evolution within GCN/BATSE trigger 7172
GRB981022 (a classical GRB) and just the 7171 GRB981022
(associated to SGR1900+14). This cornerstone link between GRB and
SGR has been finally recognized by Woods et al. (Gogus et al.
1999) very recently. Nature would be quite perverse to mimic two
very comparable events at the same detector, the same day, by the
same energy spectra and by a comparable time structures by two
totally different processes: a magnetar versus Jet GRBs. We argue
here that, apart of the energetic, both of them are blazing of
powerful jets (NS or BH); the jet are spinning and precessing
source in either binary or in accreting disk systems. The optical
transient OT  of GRB is in part due to the coeval SN-like
explosive birth of the jet related to its maximal intensity; the
OT is absent in older relic Gamma jets, the SGRs. Their explosive
memory  is left around their relic nebula or plerion  injected by
the  Gamma Jet which is running away. The late GRB OT,days after
the burst, are related to the explosion intensity; it is enhanced
only by a partial beaming $(\Omega \simeq 10^{-2} )$. The extreme
peak OT during GRB990123 (at a million time a Supernova
luminosity) is just the extreme beamed $(\Omega \leq 10^{-5} )$
Inverse Compton optical tail, responsible of the same extreme
gamma (MeV) extreme beamed $(\Omega \leq 10^{-8})$ signal.
Moreover the huge energy bath (for a fireball model) on GRB990123
imply also a corresponding neutrino burst. As in hot universe, if
entropy conservation holds, the energy density factor to be added
to the photon $\gamma$ GRB990123 budget is at least $( \simeq
(21/8)\times (4 /11)^{4/3} )$. If entropy conservation do not hold
the energy needed is at least a factor $[21/8]$ larger than the
gamma one. The consequent total energy-mass needed for the two
cases are respectively 3.5 and 7.2 solar masses. No fireball by NS
may coexist with it. Jet could. Finally Fireballs are unable to
explain the following key questions (Fargion 1998-1999) related to
the association GRB980425 and SN1998bw (Galama et all1998):

\begin{enumerate}
 \item Why nearest ``local'' GRB980425 in ESO 184-G82 galaxy at
 redshift $z_2 = 0.0083$ and the most far away ``cosmic'' ones as
 GRB971214 (Kulkarni et al.1998) at redshift $z_2 = 3.42$
 exhibit a huge average and peak
 intrinsic luminosity ratio?
\begin{equation}\label{eq1}
\frac{<L_{1 \gamma}>}{<L_{2 \gamma}>}  \cong  \frac{<l_{1
\gamma}>}{<l_{2 \gamma}>} \frac{z_{1 }^2}{z_{2 }^2} \cong 2 \cdot
10^5 \;\;; \left. \frac{L_{1 \gamma}}{L_{2 \gamma}} \right|_{peak}
\simeq 10^7 .
\end{equation}
Fluence ratios $E_1 / E_2$ are also extreme ($\geq 4 \cdot 10^5$).
 \item Why GRB980425 nearest event spectrum is softer than cosmic
 GRB971214 while Hubble expansion would naturally imply the opposite by a
 redshift factor $(1+z_1)\sim 4.43$?
 \item Why,  GRB980425 time structure is
 slower and smoother than cosmic one,as above contrary to Hubble law?
 \item Why we observed so many (even just the rare April one over 14
Beppo Sax optical transient event) nearby GRBs? Their probability
to occur, with respect to a cosmic redshift  $z_1 \sim 3.42$ must
be suppressed by a severe volume factor
\begin{equation}\label{eq2}
\frac{P_{1}}{P_{2}} \cong \frac{z_{1}^{3}}{z_{2}^{3}} \simeq 7
\cdot 10^{7} \;\;\;.
\end{equation}
\end{enumerate}
The above questions remain unanswered by fireball candle model.
Indeed hard defenders of fireball models either ignore the problem
or, worse, they negate the same reality of the April GRB event. A
family of new GRB fireballs are ad hoc and fine-tuned solutions.
We believed since 1993 (Fargion 1994) that spectral and time
evolution of GRB are made up blazing beam gamma jet GJ. The GJ is
born by ICS of ultrarelativistic (1 GeV-tens GeV) electrons
(pairs) on source IR, or diffused companion IR, BBR photons
(Fargion,Salis 1998). The beamed electron jet pairs will produce a
coaxial gamma jet. The simplest solution to solve the GRBs
energetic crisis (as GRB990123 whose isotropic budget requires an
energy above two solar masses) finds solution in a geometrical
enhancement by the jet thin beam. A jet angle related by a
relativistic kinematics would imply $\theta \sim
\frac{1}{\gamma_e}$, where $\gamma_e$ is found to reach $\gamma_e
\simeq 10^3 \div 10^4$ (Fargion 1994,1998). At first approximation
the gamma constrains is given by Inverse Compton relation: $<
\epsilon_\gamma > \simeq \gamma_e^2 \, k T$ for $kT \simeq
10^{-3}-10^{-1}\,eV$ and $E_e \sim GeVs$ leading to characteristic
X-$\gamma$ GRB spectra. However an impulsive unique GRB jet burst
(Wang \& Wheeler 1998) increases the apparent luminosity by
$\frac{4 \pi}{\theta^2} \sim 10^7 \div 10^9$ but face a severe
probability puzzle due to the rarity to observe even a most
frequent SN burst jet pointing in line toward us. Viceversa one
must assume a high rate of GRB events ($\geq 10^5$ a day larger
even than expected SN one a day). Most authors today are in a
compromise: they believe acceptable  only mild beaming ($\Omega
> \sim 10^{-3}$), taking GRB980425 out of the GRB ''basket''. On
the contrary we considered
 GRBs and SGRs as multi-precessing
and spinning Gamma Jets and the GRB980425 an off-axis classical
jet. In particular we considered (Fargion 1998) an unique scenario
where primordial GRB jets decaying in hundred and thousand years
become the observable nearby SGRs. Sometimes accretion binary
systems may increase the SGRs activity. The ICS for monochromatic
electrons on BBR leads to a coaxial gamma jet spectrum(Fargion \&
Salis 1995,1996,1998):
$\frac{dN_{1}}{dt_{1}\,d\epsilon_{1}\,d\Omega _{1}}$ is
\begin{equation}
\epsilon _{1}\ln \left[ \frac{1-\exp \left( \frac{-\epsilon
_{1}(1-\beta \cos \theta _{1})}{k_{B}\,T\,(1-\beta )}\right)
}{1-\exp \left( \frac{-\epsilon _{1}(1-\beta \cos \theta
_{1})}{k_{B}\,T\,(1+\beta )}\right) }\right] \left[ 1+\left(
\frac{\cos \theta _{1}-\beta }{1-\beta \cos \theta _{1}}\right)
^{2}\right] \label{eq3}
\end{equation}
scaled by a proportional factor $A_1$ related to the electron jet
intensity. The adimensional photon number rate (Fargion \& Salis
1996) as a function of the observational angle $\theta_1$
responsible for peak luminosity (eq. \ref{eq1}) becomes
\begin{equation}
\frac{\left( \frac{dN_{1}}{dt_{1}\,d\theta _{1}}\right) _{\theta
_{1}(t)}}{ \left( \frac{dN_{1}}{dt_{1}\,d\theta _{1}}\right)
_{\theta _{1}=0}}\simeq \frac{1+\gamma ^{4}\,\theta
_{1}^{4}(t)}{[1+\gamma ^{2}\,\theta _{1}^{2}(t)]^{4}}\,\theta
_{1}\approx \frac{1}{(\theta _{1})^{3}} \;\;.\label{eq4}
\end{equation}
The total fluence at minimal impact angle $\theta_{1 m}$
responsible for the average luminosity (eq. \ref{eq1}) is
\begin{equation}
\frac{dN_{1}}{dt_{1}}(\theta _{1m})\simeq \int_{\theta
_{1m}}^{\infty }\frac{ 1+\gamma ^{4}\,\theta _{1}^{4}}{[1+\gamma
^{2}\,\theta _{1}^{2}]^{4}} \,\theta _{1}\,d\theta _{1}\simeq
\frac{1}{(\,\theta _{1m})^{2}}\;\;\;. \label{eq5}
\end{equation}
These spectra fit GRBs observed ones (Fargion \& Salis 1995).
Assuming a beam jet intensity $I_1$ comparable with maximal SN
luminosity, $I_1 \simeq 10^{45}\;erg\,s^{-1}$, and replacing this
value in adimensional $A_1$ in equation \ref{eq3} we find a
maximal apparent GRB power for beaming angles $10^{-3} \div
3\times 10^{-5}$, $P \simeq 4 \pi I_1 \theta^{-2} \simeq 10^{52}
\div 10^{55} erg \,s^{-1}$ within observed ones. We also assume a
power law jet time decay as follows
\begin{equation}\label{eq6}
  I_{jet} = I_1 \left(\frac{t}{t_0} \right)^{-\alpha} \simeq
  10^{45} \left(\frac{t}{3 \cdot 10^4 s} \right)^{-1} \; erg \,
  s^{-1}
\end{equation}
where ($\alpha \simeq 1$) able to reach, at 1000 years time
scales, the present known galactic microjet (as SS433) intensities
powers: $I_{jet} \simeq 10^{38}\;erg\,s^{-1}$. We used the model
to evaluate if April precessing jet might hit us once again. It
should be noted that a steady angular velocity would imply an
intensity variability ($I \sim \theta^{-2} \sim t^{-2}$)
corresponding to some of the earliest afterglow decay law.

\section{The GRB980425-GRB980712 repeater}

Therefore the key answers to the above puzzles (1-4)  are: the
GRB980425 has been observed off-axis by a cone angle wider than
$\frac{1}{\gamma}$ thin jet  by a factor $a_2 \sim 500$ (Fargion
1998) $\theta \sim \frac{500}{10^4} \approx \frac{5 \cdot
57^0}{100} \approx 2.85^0 \left( \frac{\gamma}{10^4}
\right)^{-1}$, and therefore one observed only the ``softer'' cone
jet tail whose spectrum is softer and whose time structure is
slower (larger impact parameter angle). A simple statistics
favoured a repeater hit. Indeed GRB980430 trigger 6715 was within
$4 \sigma$ and particularly in GRB980712 trigger 6917 was within
$1.6 \sigma$ angle away from the April event direction. An
additional event 15 hours later, trigger 6918, repeated making the
combined probability to occur quite rare ($\leq 10^{-3}$). Because
the July event has been sharper in times ($\sim 4 \,s $) than the
April one ($\sim 20 \,s $), the July impact angle had a smaller
factor $a_3 \simeq 100$. This value is well compatible with the
 expected peak-average luminosity flux evolution in eq.(6,4):
$\frac{L_{04\,\gamma}}{L_{07\,\gamma}} \simeq
\frac{I_2\,\theta_2^{-3}}{ I_3\,\theta_3^{-3}} \simeq \left(
\frac{t_3}{t_2} \right)^{-\alpha} \,\left( \frac{a_2}{a_3}
\right)^{\,3} \leq 3.5$ where $t_3 \sim 78 \; day$ while $t_2 \sim
2 \cdot 10^5 \, s$. The predicted fluence is also comparable with
the observed ones $\frac{N_{04}}{N_{07}} \simeq
\frac{<L_{04\,\gamma}>}{<L_{07\,\gamma}>} \, \frac{\Delta
\tau_{04}}{\Delta \tau_{07}} \simeq \left( \frac{t_3}{t_2}
\right)^{-\alpha} \,\left( \frac{a_2}{a_3} \right)^2
\,\frac{\Delta \tau_{04} }{\Delta \tau_{07}} \geq 3$.
\section{The SGRs hard spectra and their GRB link}
Last SGR1900+14 (May-August 1998) events and SGR1627-41
(June-October 1998) events did exhibit at peak intensities hard
spectra comparable with classical GRBs. We imagine their nature as
the late stages of jets fueled by a disk or a companion (WD,NS)
star. Their binary angular velocity $\omega_b$ reflects the beam
evolution $\theta_1(t) = \sqrt{\theta_{1 m}^2 + (\omega_b t)^2}$
or more generally a multi-precessing angle $\theta_1(t)$ (Fargion
\& Salis 1996) which keeps memory of the pulsar jet spin
($\omega_{psr}$), precession by the binary $\omega_b$ and
additional nutation due to inertial momentum anisotropies or
beam-accretion disk torques ($\omega_N$). On average, from eq.(5)
the gamma and afterglow decays as $t^{-2}$; the complicated
spinning and precessing jet blazing is responsible for the wide
morphology of GRBs and SGRs as well as their internal periodicity.
In conclusion the puzzles for GRB980425-GRB971214 find a simple
solution within a precessing jet:
 The different geometrical
observational angle might compensate the April 1998 low peak gamma
luminosity ($10^{-7}$) by a larger impact angle which compensates,
at the same time, its statistical rarity ($\sim 10^{-7}$) its
puzzling softer nature and longer timescales. Such precessing jets
may explain (Fargion \& Salis 1995) the external twin rings around
SN1987A. We predicted its relic jet to be found in the South-East
due to off-axis beaming acceleration. Jets may propel and inflate
plerions as the observed ones near SRG1647-21 and SRG1806-20.
Optical nebula NGC6543 (``Cat Eye'') and its thin jets fingers (as
Eta Carina ones), the  double cones sections in Egg Nebula CRL2688
are the most detailed and spectacular lateral view of such jets
''alive''. Their blazing in-axis would appear in our galaxy as
SGRs or, at maximal power at their SN birth at cosmic edges, as
GRBs.

\section{The Morphology of Precessing Jet  relics}
  The Gamma Jet progenitor of the GRB
   is leaving a trace in the space : usually a nebulae
where  the nearby  ISM  may record the jet  sweeping as a
     three dimensional screen. The outcomes  maybe either
      a twin ring as recent SN1987A has shown, or helix traces as
       the Cat Eye Nebula or more structured shapes as plerions
        and hourglass nebulae.
        How can we explain within an unique jet model such a wide
         diversity?\\

         We imagine the jet as born by a binary system (or
         by an asymmetric disk accreting interaction)
         where the compact companion (Bh or NS) is the source of the
         ultra relativistic electron pair jet (at tens GeV.
       Inverse Compton Scattering on IR thermal photons will produce a collinear
        gamma jet at MeV). The rarest case where the jet
         is spinning and nearly isolated would produce a jet train
         whose trace are star chains as the Herbig Haro ones
         (Fargion, Salis 1995). When the jet
         is modified  by the magnetic field torque of the
         binary companion field the result may be a more rich cone shape.
         If the ecliptic  lay on the same plane orthogonal to the jet
         in an ideal circular orbit than the bending
         will produce an ideal twin precessing cones which is
           reflected in an ideal twin rings (Fargion,Salis 1995).
         If the companion is in eccentric orbit the resultant
         conical jet will be more deflected at perihelion while
          remain nearly undeflected at a aphelion.
          The consequent off-axis cones will play the role of a
          mild "rowing" acceleration  able to move the system
          and speed it far from its original birth (explosive)
          place. Possible traces are the asymmetric external twin
          rings painted onto the spherical relic shell by SN1987a.
          Fast relics NS may be speeded by this processes (Fargion, Salis
          1995a, 1995b, 1995c). Because of momentum conservation this asymmetric
            rowing is the source of a motion of the jet relic  in the
            South-East direction. In extreme eccentric system
            the internal region of the ring are more  powered
            by the nearby encounter leading to the apparent gas arcs.
            If the system is orbiting in a plane different from
              the one orthogonal to the jet the outcoming
              precessing jet may spread into a mobile twin cone
              whose filling may appear as a full cone or a twin hourglass
              by
               a common plerion shape. At late times there is also  possible
               apparent spherical shapes sprayed and structured by a chaotic
               helix.
               External  ISM distribution may also play a role enhancing
                 some sides or regions of the arcs.
              The integral jet in  long times may mimic even
                spherical envelopes but internal detailed inspection
                might reveal the thin jet origin (as in recent Eta
                Carina string jets). Variable nebulae behaviours recently
                observed are confirming our present scenario.

\section{Would SS433-like objects trigger star formation?}

   There are example objects with energetic relativistic jets in our own
Galaxy - the microquasars and the object SS433. The existence of
such mini-precessing jets was the first starting point (Fargion
1994) in understanding the jets blazing role in SGR-GRB. The
recent HI data for the latter did prove that the accumulated
energy of its jet is large enough (Dubner et al., 1998).  These
authors found the HI shell around SS433 with the velocity of
expansion of about 76 km/s and the radius 1.1 degree (56 pc if the
distance is 3 kpc) and evaluated the kinetic energy transferred to
the surrounding medium to be $\sim 2 \cdot 10^{51}$ erg. They
concluded that the input of kinetic energy from the jets of SS433
being $\sim 10^{39}$ ergs/s (Margon 1984) and the life-time being
$\sim 2 \cdot 10^4$ yr (Zealey et al. 1980), the total energy
injected by jets is $\sim 10^{51}$ ergs. This is just compatible
with energy which was necessary to swept-up the gas and produce
the LMC4 arcs (Efremov and Elmegreen 1998). Moreover, this jet may
well last still longer. The total mass of HI which was swepted-up
by SS433 is $\sim$ 30 000 $M_{\odot}$. It is quite possible that
later on the star formation will start there. Anyway, there is the
suggestion that SS433 is the precursor of a GRB (Pugliese et al.
1999).

Objects of the Galaxy known as microquasars have also powerful
relativistic jets, which  may have kinetic energy reaching
$10^{43}$ ergs; there are some indication of induced star
formation at the jet ends (Rodriguez and Mirabel, 1998) These
objects are surely binary stars with compact component and X-ray
sources; 9 such events are known in the Galaxy and thus are
presumed relatives to the  GRBs and SGRs. These long acting jets
may well be the trigger of star formation in  the pre-existing or
in the swepted-up gas, and the stellar arcs may indicate the
latter case.

The analogous star formation  induced by jets from the active
galactic nuclei is well known phenomenon. The  nearest example of
such event is in Cen A galaxy (Graham, 1998). One of us (DF)
believes that similar precessing jets links AGN to GRB and SGRs.

\section{Conclusions}

The concentration of the giant stellar arcs and plausible GRB
progenitors (or at least their close relatives) to the same and
the only region of the LMC, the same in which the only (in the LMC)
SGR is located, strongly suggests that GRB and the arcs are  genetically
connected phenomena.  The giant arcs  are plausibly the stellar
remnants of GRB events.   One way to produce these events is
merging in compact binary systems and there is suggestion that
ejections of such binaries from the dense old cluster NGC 1978
explain the concentration of all these objects in the LMC4 region.
At any rate, there should  be the common source of the multiple
arcs nearby. The  hardening (i.e. shortening the orbital period)
leading to the heating the stars and the final ejection of close binary
stars from a dense cluster is a process
generally assumed in studies of the dynamical evolution of star clusters.
The close binaries of two compact  objects are formed in a dense  cluster
and seems to be unavoidable ejected from it during a few Billions years, before
getting hardening sufficient enough for the subsequent merging, which
need at least millions year more in emission of the gravitational waves
(Portegiez Zwart and McMillan, 2000). It is therefore possible to suggest
that X-ray emission and more so the GRB event (in the merging of two compact
objects) can arise  only after ejection, thus explaining
the absence of X-ray source in NGC 1978 itself, as well as the absence
of the stellar arc centered on the cluster (Efremov 1999, 2000).

The orientation of apexes of stellar arcs in the LMC  suggests
that the arc-like shapes are not due to the outburst outside the middle
of the gas disk, i.e. the vertical density gradient (Efremov et al. 1999).
The stellar arcs in  the LMC, in NGC 6946 and other galaxies  are
always  perfectly round and the only explanation is that they are
parts of spherical shells. These  two features  could be explained
with two possibilities: the events which triggered  star formation
in the partial shells were either   the collimated (at the opening
angle about 60 - 90 degree) super-explosions  or these shells
resulted from the narrow precessing jets, the precession angle
being  variable and its maximal value being 60 - 90 degree. As the
data on the HI shell around SS433 and also on the mini-quasars in the
Galaxy demonstrate, the long standing stellar jets may form a gas
shell large enough. The regions of star formation, triggered in
such an way  may be numerous enough, because depending on the
density fluctuations of the surrounding gas and especially on the
projection angles, they might not look like the stellar arcs (Efremov,
1999, 2000).

If the SGR, three or four stellar arcs and supershell LMC4 in the
LMC are really the imprints of GRB, we have five or six GRB events
or peak activities of a slow motion SGR  in the LMC during last 30
or so Myrs. This seems to be in contradiction with the high
frequency of GRB-SN events suggested by the possibility that they
are narrow jets, yet it is possible that most of these arise far
from the gas disk of a galaxy (merging in the binary with compact
components may occurs in Billions years after its formation  in
the galaxy disk and their velocity may be as high as 100 km/s or
so, e.g. Lipunov et al. 1997) or at least in the low density areas
where the probability for a jet to meet the gas cloud is low,
leading to negligible gas accumulation and small star formation.

The existence of the stellar arcs give the strong support to the
possibility that the GRB as well as SGR events are connected with
the multi-precessing narrow jets, first suggested by Fargion
(1994, 1995, 1998, 1999).The Gamma Jet energetic , triggered by an
inner Comptonizing electron pairs jet, lead to GRBs spectra
(Fargion,Salis 1995-1998)and  observed variable morphology. New
evidences favor the
 same SN1987A external twin rings  as been relics of a precessing
 or multi-precessing jet. There are wider consensus today on the GRB-SN-Jet connection
(Blackman 1996;Wang,Wheeler;1998; Dar 1999) and the various
features of the Gamma-ray light curves during the GRB events were
modelled recently with the multiprecessing jets (Portegeis Zwart
et al., 1999).

The surprising connection between SN-GRB and stellar arc
origination favor recent evidence for GRB preferentially located
in the isolated star formation regions.  This regions might be triggered by
SN-GRB  events,  progenitors of which might have had the common
origin in a massive cluster near the region.

The key problems of GRB and SGR as well as stellar arcs seem to be
  somehow linked together and solved at once. However  main open
  questions are still puzzling:  what are the real physical processes
   leading to such high,  laser like, collimating and powerful, SN like,
    astrophysical jets? Do these jet contain  also a hard spectra tail
    extending up to ultra high GZK energy frontiers (Fargion,Mele,Salis 1999)?
\\

\ {Figure 1 Caption}
\\

The four giant stellar arcs in the  region of the supershell LMC4
in the LMC.  The arcs  have the perfect circular shape, especially
evident for the smallest and youngest ones. The old massive
cluster NGC 1978 is surrounded by a  circle.

\end{document}